\documentclass[twocolumn,prb,aps,superscriptaddress]{revtex4}
\usepackage{color}
\usepackage{graphicx}
\usepackage{subfig}
\usepackage{natbib}
\usepackage[fleqn]{amsmath}
\usepackage{amssymb}
\newcommand{\nc}{\newcommand}
\nc{\on}{\operatorname}
\nc{\wt}{\widetilde}
\nc{\Wick}{{\mathbb :}}
\nc{\R}{{\mathbb R}}

\newcommand{\beq}{\begin{equation}}
\newcommand{\eeq}{\end{equation}}
\newcommand{\bmul}{\begin{multline}}
\newcommand{\emul}{{\end{multline}}}
\newcommand\beqa{\begin{eqnarray}}
\newcommand\eeqa{\end{eqnarray}}
\newcommand\bea{\begin{array}}
\newcommand\eea{\end{array}}
\newcommand\ba{\begin{array}}
\newcommand\ea{\end{array}}

\newcommand{\neqa}{\nonumber\end{eqnarray}}

\nc{\CH}{{\mathcal H}}
\nc{\Db}{{\bar D}}
\nc\comment[1]{}

\nc{\CM}{{\mathcal M}}
\nc{\CN}{{\mathcal N}}

\newcommand{\re}{\relax{\rm I\kern-.18em R}}

\nc{\meV}{{\mathrm{\,meV}}}
\nc{\cG}{{\mathcal G}}

\renewcommand{\)}{\right)}

\comment{ \)}
\renewcommand{\bar}{\overline}

\nc{\al}{{\alpha}}

\begin{document}

\title{Bogoliubov-Born-Green-Kirkwood-Yvon Chain and Kinetic Equations for the Level Dynamics in an Externally Perturbed Quantum System}
\author{Mumnuna A. Qureshi}
\address{Department of Physics, Loughborough University, Loughborough LE11 3TU, UK.}
\author{Johnny Zhong}
\address{Department of Mathematical Sciences, Loughborough University, Loughborough LE11 3TU, UK.}
\author{Joseph J. Betouras }
\address{Department of Physics, Loughborough University, Loughborough LE11 3TU, UK.}
\author{Alexandre M. Zagoskin}
\address{Department of Physics, Loughborough University, Loughborough LE11 3TU, UK.}

\begin{abstract}
Theoretical description and simulation of large quantum coherent systems out of equilibrium remains a daunting task. Here we are developing a new approach to it based on the Pechukas-Yukawa formalism, which is especially convenient in the case of an adiabatically slow external perturbation though not restircted to adiabatic systems. In this formalism the dynamics of energy levels in an externally perturbed quantum system as a function of the perturbation parameter is mapped on that of a fictitious one-dimensional {\em classical} gas of particles with repulsive potential. Equilibrium statistical mechanics of this Pechukas gas allows to reproduce the random matrix theory of energy levels. In the present work, we develop the non-equilibrium statistical mechanics of the Pechukas gas, starting with the derivation of the Bogoliubov-Born-Green-Kirkwood-Yvon (BBGKY) chain of equations for the appropriate generalized distribution functions. Sets of approximate kinetic equations can be consistently obtained by breaking this chain at a particular point (i.e. approximating all higher-order distribution functions by the products of the lower-order ones). When complemented by the equations for the level occupation numbers and inter-level transition amplitudes, they allow to describe the non-equilibrium evolution of the quantum state of the system, which can describe better a large quantum coherent system than the currently used approaches. In particular, we find that corrections to the factorized approximation of the distribution function scale as 1/N, where N is the number of the "Pechukas gas particles" (i.e., energy levels in the system).
\end{abstract}
\maketitle

\section{Introduction}

The study of quantum many body systems out of equilibrium has attracted enormous attention in the recent years, especially in application to artificial quantum coherent structures. One of the main stimuli for this research is the quest for a practically useful quantum computer. Assuming the universal applicability of quantum mechanics, the practical realization of a universal quantum computer remains a rather distant possibility due to the large number of physical qubits necessary for its operation and the extreme fragility of its quantum states with respect to external and internal sources of decoherence  \cite{Zagoskin1, Requist4, Fahri1}. Moreover, as was shown by Feynman \cite{Feynman}, simulation of a large enough quantum coherent system by classical means is impossible due to the dimensionality of the corresponding Hilbert space growing exponentially with the size of the system (e.g., number of qubits). Unfortunately, the size of a practically useful universal quantum computer greatly exceeds the limits of what is tractable by classical means. This makes the task of determining the degree of "quantumness" of such a device, its design and optimization exceedingly difficult \cite{Zagoskin1, Requist4, Fahri1, Fahri2, Sarovar, Candia}. An attractive alternative is adiabatic quantum computing (AQC) \cite{Fahri1, Fahri2}
The starting point is to consider a system governed by the general Hamiltonian \cite{Pechukas,Yukawa3,Haake}:
\begin{equation}
H(\lambda(t))=H_0+\lambda(t)ZH_b,
\end{equation}

 \noindent where $H_0$ is a complex unperturbed Hamiltonian with an easily achievable nondegenerate ground state, 
 $\lambda(t)$ is a time evolving parameter generally taken to be adiabatic, $ZH_b$ is a large bias perturbation term with $Z\gg1$. A significant progress was achieved in theoretical and experimental research in this direction \cite{Requist4, Pechukas, Yukawa1, Yukawa2,Yukawa3, Haake, Yvon,  Berends, Huyghebaert, Poulin}. Nevertheless the development of adiabatic quantum computers faces the same fundamental problem of impossibility of their direct simulation by classical means  \cite{Zagoskin1, Requist4, Fahri1}. This stimulates the search for alternative theoretical methods, which could provide some useful figures-of-merit describing large quantum systems out of equilibrium. A common approach to achieve quantum coherence in non-equilibrium many body dynamics is Keldysh Green function theory\cite{Requist1}, however this approximation is limited to short time intervals where its errors grow as a power of time\cite{Requist1}. This paper aims at developing basic elements of such an approach, which would seem especially useful for, but not necessarily restricted to,  modelling of adiabatic quantum computers. It was shown in Ref. \onlinecite{Zagoskin} that the parametric evolution of the system described by Eq. (1) can be mapped on the classical Hamiltonian dynamics of a one-dimensional (1D) gas model with long-range repulsion; the Pechukas gas. The equilibrium statistical mechanics of Pechukas-Yukawa gas turned out a useful tool in justifying the Random Matrix Theory \cite{Haake}.
In Ref.(\onlinecite{Zagoskin}) this approach was successfully used to describe the operation of a small-scale adiabatic quantum computer, but its scaling up was restricted for the same reason as mentioned above, and it was suggested that building the kinetic theory of the Pechukas-Yukawa gas may provide a useful solution.

In this work, we develop a consistent description of a non-equilibrium, nonstationary evolution of a perturbed quantum system based on the kinetic theory of Pechukas-Yukawa gas. The formalism is applicable to an arbitrary system described by a Hamiltonian of the form of Eq. (1). As in classical kinetic theory, we expect that the statistical approach to the level dynamics (as functions of the parameter $\lambda$) would allow a reduced description in terms of correlation functions, which can be used as a basis for controlled approximations. Adding to this scheme the time evolution of the system's quantum state, we will be able to build an expansion in terms of the adiabatic parameter, $\dot{\lambda}$. This could provide a better insight into what measurable characteristics of a system can be used as criteria of its quantum performance, and also make possible an approximate simulation of larger systems than those tractable by other methods. We test the theory for a small system of two interacting qubits, simulated by a transverse field Ising Hamiltonian (TFIH). Furthermore, the relationship between the level dynamics and that of the occupation numbers as a function of time is established using the Pechukas model and we obtain the full wave function that describes the system. 

As a result, the application of the BBGKY hierarchy to the Pechukas model, parametrically driven evolution of a quantum system, is especially useful in accomodating for adiabatic systems, however the formalism is applicable to a general system with parametric evolution in time, exploring an important new direction in contemporary physics and open further investigations in order to understand the connection to the physics of the Pechukas gas. This description is advantageous as it is expected to have significant developments in non-equilibrium processes such as decohenrence\cite{Zagoskin1, Fahri1, Requist1}. We, eventually, would like to address the question to what extent this approach can describe adiabatic quantum computing. 

The structure of the paper is such that in Sec. II we give a brief overview of the Pechukas equations and the mapping to the interacting gas model, while in Sec III a brief overview of BBGKY hierarchy is provided. In Sec. IV we present the BBGKY hierarchy for the Pechukas model and discuss the factorisation approximation. In Sec. V we estimate the relative error ${\cal{E}}_r$ from the factorisation approximation. Sec. VI is devoted to a two-qubit system governed by TFIH. In Sec VII we present the equations of the evolution of occupation numbers and the density matrix and we conclude in Sec VIII. For clarity, some long derivations are relegated to the Appendices.

\section{Pechukas Equations}
The Hamiltonian Eq. (1) that governs a complex system that evolves in time parametrically through $\lambda$, generally taken to be adiabatic, is described through the instantaneous eigenstates $|m(\lambda)>$ and eigenvalues $E_m (\lambda)$ which are related in the following way:

\begin{equation}
H(\lambda)|m\rangle =E_m (\lambda).
\end{equation}

\noindent The contribution from $H_0$ is fully determined at all times and information of all initial conditions are obtained through instantaneous matrix elements of the following form:

\begin{equation}
\langle m|H_0 |n\rangle = E_m (\lambda) \delta_{mn}-\lambda \langle m|ZH_b |n\rangle.
\end{equation}

\noindent The contribution from the parametrically evolved term $\lambda \left(t\right)Z{\mathcal{H}}_b$, is determined through the Pechukas equations \cite{Pechukas, Yukawa1, Yukawa2,Yukawa3, Haake, Yvon, Balescu, Zagoskin, Wilson1, Wilson2}. Taking the Hamiltonian that describes the Pechukas gas as $H(\lambda(t))=\frac{1}{2}\sum_{m}v^2_{m}+\frac{1}{2}\sum_{m \neq n}\frac{|l_{mn}|^2}{(x_m-x_n)^2}$, with unit mass, one can derive a closed set of first order ordinary differential equations that describe the ``position'' ($x_m$), ``velocity'' ($v_m$) and ``relative angular momentum'' ($l_{mn}$), as expressed in Eq. (4). These are obtained by taking the Poisson brackets of these coordinates with the Hamiltonian to give the source-free Hamiltonian flow \cite{Zagoskin, Haake}.

\begin{equation}
\begin{gathered}
\dot{x}_m=v_m \\
\dot{v}_m=2\sum_{m\neq n}{\frac{{{|l}_{mn}|}^2}{{(x_m-x_n)}^3}}\\
\dot{l}_{mn}=\sum_{k\neq m,n}{l_{mk}l_{kn}\left(\frac{1}{{(x_m-x_k)}^2}-\frac{1}{({x_k-x_n)}^2}\right)},
\end{gathered}
\end{equation}

\noindent where $x_m\left(\lambda \right)=E_m(\lambda )$, $v_m\left(\lambda \right)=\left\langle m|Z{\mathcal{H}}_b|m\right\rangle $ and $l_{mn}\left(\lambda \right)=\left(E_m\left(\lambda \right)-E_n(\lambda )\right)\left\langle m|Z{\mathcal{H}}_b|n\right\rangle $ which is an antisymmetric complex quantity as $l_{mn}=-l^*_{nm}$. The indices ($m$) represent the positions, velocities and particle-particle repulsion as determined by the relative angular momenta, for the corresponding $m^{th}$ particle interaction. Here $\lambda $ plays the role of time \cite{Pechukas, Yukawa1, Yukawa2,Yukawa3,Haake, Zagoskin}. This procedure describes the aforementioned mapping of the level dynamics of a system to that of a one-dimensional classical gas. It is worth stressing that the mapping of Eq. (1) on Eq. (4) is an identical operation valid for an arbitrary choice of $H_0$ and $H_b$ and an arbitrary time dependence of $\lambda$, not necessarily adiabatic. Note that time does not explicitly enter Eq. (4), rather concerning the evolution in time parametrically through $\lambda$ which determines the instantanoues energy levels: this is a set of equations for the Hamiltonian, and not for (time- and initial state-dependent) quantum states of a system described by such a Hamiltonian.

 The Pechukas equations Eq. (4), govern the evolution of a system of particles moving in one dimension of which interact through long range coupling described by the relative angular momenta\cite{Zagoskin, Haake}. Without loss of generality, phase effects are taken to be zero where the usual Berry phase relating to adiabatic systems is not present here as the system is not cyclic\cite{Haake, Requist1}. Using this model, one can determine the dynamics of microscopic distribution functions following standard kinetic approaches involving the BBGKY hierarchy. These are derived in the next section. 

\section{BBGKY Hierarchy}
The standard BBGKY hierarchy arises from the continuity equation in phase space. It is used to describe the time evolution of classical reduced distribution functions for a general Hamiltonian. The chain relates the reduced distribution function for $s$ particles to the distribution function for $s+1$ particles, where $s \in {[1,2, \dots N]}$ concerning positions and velocities \cite{Yvon, Balescu, Stockmann, Alexeev, Requist1, Suzuki}.

We denote an empirical distribution function, $F_N(x_1\dots x_N,v_1\dots v_N)$ averaged over the initial conditions, of the form: 

\begin{equation}
\begin{gathered}
F_N(x_1\dots x_N,v_1\dots v_N)=\\
\langle \prod_m{\delta \left(x_m-{\xi }_m\right)\delta (v_m-{\omega }_m)}\rangle.
\end{gathered}
\end{equation}

\noindent The averaging procedure is described through:
\begin{equation}
\langle f (x_m,v_n )\rangle := \frac{1}{|I|} \sum_{x^0,\ v^0\in I} 〖f (x^t,v^t;x^0,v^0)〗,
\end{equation}

\noindent where $\left|I\right|$ denotes the size of $I$, the set of initial conditions. And $f(x^t,v^t;x^0,\ v^0)$ denotes the function evaluated at $(x^t,v^t,t)$, the propagated coordinates up to time $t$, parameterised by $(x^0,\ v^0)$. This is essentially a counting function of where the particles are present.

The BBGKY hierarchy for this arbitrary distribution function $F_N\left(x_1\dots x_N,v_1\dots v_N\right),$ in phase space is defined through the following set of equations:
\begin{equation}
\begin{gathered}
\partial_tF_s=
{} \sum^s_{j=1}{L^0_j}F_s
+ \sum^s_{j=1}{L^F_j}F_s
+ \sum^s_{j=1}{\sum^{j-1}_{n=1}{L^I_{jn}}}F_s\\
+ \sum^s_{j=1}{{\int{dx_{s+1}}L^I_{j(s+1)}}F_{s+1}}.
\end{gathered}
\end{equation}

The reduced distribution function, $F_s := F_s(x_1\dots x_s,v_1\dots v_s)$ taken up to the $s$-particle interactions hence it takes into account only the distribution functions of the $s$-particle and the $(s+1)$-particle. The first term, $L^0$ corresponds to the free part of the Hamiltonian, the second term, $L^F$ describes the external field, e.g. noise in the system. The final two terms associated with $L^I$ correspond to the perturbation contribution of the Hamiltonian as result of interactions \cite{Yvon, Balescu, Stockmann, Alexeev, Requist1, Requist2, Requist3, Sergey}. 





 Although this hierarchy produces a scheme which determines the kinetic equations of motion, it does not accommodate for the nature of quantum systems with parametrically evolving Hamiltonians\cite{Requist1, Meyer, Stock}. In the following we derive a generalised BBGKY hierarchy that describes the level dynamics associated to non-equilibrium systems which extends to parametrically evolving Hamiltonians, using the Pechukas model. 

\section{BBGKY Hierarchy for the Pechukas Model and Factorisation Approximation}
\noindent We derive the BBGKY hierarchy for the Pechukas model concerning level dynamics with respect to a full distribution of fictitious  ``position", ``velocity" and ``relative angular momentum", evolving parametrically with time through $\lambda$. Application of the BBGKY hierarchy to parametrically driven evolution of a quantum system is particularly useful in describing non-equilibrium adiabatic systems, taking $\lambda$ to be an adiabatic parameter,  such that the kinetic equations of the level dynamics are obtained as given by the following equation.  However the formalism is general, applicable to all parametrically evolving systems. We relegate the details of the derivation to Appendix A and present here the final scheme:

\begin{widetext}
\begin{eqnarray}
\nonumber
\frac{\partial }{\partial \lambda}F_{s,s(s-1)}
=\sum^s_{m=1}{v_m\frac{\partial }{\partial x_m}F_{s,s(s-1)}}
\nonumber
+2\sum^s_{m=1}{\sum^{m-1}_{n=1}{(\frac{{{|l}_{mn}|}^2}{{(x_m-x_n)}^3}+\frac{{{|l}_{nm}|}^2}{{(x_n-x_m)}^3})\frac{\partial }{\partial v_m}F_{s,s(s-1)}}}\\
\nonumber
+2\int{dx_{s+1}dv_{s+1}\boldsymbol{D}{\boldsymbol{l}}_{s+1}}
\sum^s_{m=1}{(\frac{{{|l}_{m(s+1)}|}^2}{{(x_m-x_{s+1})}^3}+\frac{{{|l}_{(s+1)m}|}^2}{{(x_{s+1}-x_m)}^3})\frac{\partial }{\partial v_m}F_{s+1,s\left(s+1\right)}}\\
\nonumber
+\sum^s_{m=1}{\sum^{m-1}_{k=1}{\sum^{k-1}_{n=1}{l_{mk}l_{kn}(\frac{1}{{(x_m-x_k)}^2}-\frac{1}{({x_k-x_n)}^2})\frac{\partial }{\partial l_{mn}}}F_{s,s\left(s-1\right)}}}\\
+\int{dx_{s+1}dv_{s+1}\boldsymbol{D}{\boldsymbol{l}}_{s+1}}\sum^s_{k=1}{\sum^{k-1}_{n=1}{l_{s+1k}l_{kn}(\frac{1}{{(x_{s+1}-x_k)}^2}-\frac{1}{({x_k-x_n)}^2})\frac{\partial }{\partial l_{(s+1)n}}}F_{s+1,s\left(s+1\right)}},
\end{eqnarray}

\noindent where $F_{N,N(N-1)} = F_{N,N(N-1)}(x_1,\dots x_N, v_1,\dots v_N, l_{12}, \dots l_{N,N-1})$ and $F_{s,s(s-1)}=F_{s,s(s-1)}(x_1,\dots x_s, v_1,\dots v_s, l_{12}, \dots l_{s,s-1})$ describing the reduced distribution function up to $s$-particle interactions and $F_{s+1,s\left(s+1\right)}=F_{s+1,s\left(s+1\right)}(x_1, \dots x_{s+1},v_1 \dots v_{s+1},l_{12},\dots l_{s+1,s})$ is the reduced distribution function up to the $(s+1)$ particle interactions. We denote $\boldsymbol{D}{\boldsymbol{l}}_{s+1}\prod^s_{i=1}{dl_{s+1,i}dl_{i,s+1}}$.
\end{widetext}

The reduced distribution functions of the energy levels reflect the average density of levels so that, by deriving the BBGKY chain, we describe the evolution in parameter $\lambda$, of the level dynamics \cite{Haake}. This is useful as there is a direct relationship between the level dynamics of the Pechukas system and that of its quantum states as seen in Section VII. Here, $F_{s,s(s-1)}$ is defined by the following ($F_{s+1,s\left(s+1\right)}$ is defined similarly):

\begin{widetext}
\begin{equation}
F_{s,s\left(s-1\right)}{{:=}}
\frac{N!}{(N-s)!}.\frac{(N^2-N)!}{(N^2-N-s(s-1)!}
\int{dx_{s+1}\dots dx_ndv_{s+1}\dots dv_ndl_{s+1,s}\dots dl_{n,n(n-1)}F_{N,N(N-1)}}.
\end{equation}
\end{widetext}

\noindent With probabiliy distribution, $F_{N,N(N-1)}$ and dynamic variables $x_m,v_n$ and $l_{mn}$,  averaging over $\xi ,\ \omega ,\ \Lambda$ defined in Appendix A.



To illustrate the scheme more explicitly we write the BBGKY chain up to the 2${}^{nd}$ equation. We neglect the $s=0$ level as this merely vanishes on the right hand side of the chain. Starting from $s=1$ we obtain that $F_{1,0}(x_1,\ v_1)$ the associated chain is:

\begin{widetext}
\begin{equation}
\begin{gathered}
\frac{\partial }{\partial \lambda}F_{1,0}(x_1,v_1)=v_1\frac{\partial }{\partial x_1}F_{1,0}(x_1,v_1)\\
+2\int{dx_2dv_2dl_{12}dl_{21}}(\frac{{{|l}_{12}|}^2}{{(x_1-x_2)}^3}+\frac{{{|l}_{21}|}^2}{{(x_2-x_1)}^3})\frac{\partial }{\partial v_1}F_{2,2}(x_1,x_2,v_1,\ v_2,l_{12},l_{21}). 
\end{gathered}
\end{equation}
\end{widetext}

In the same way, the chain has been explicitly constructed for $s=2$ with $F_{2,2}\left(x_1,x_2,v_1,\ v_2,l_{12,\ \ }l_{21}\right)$ :

\begin{widetext}
\begin{equation}
\begin{gathered}
\frac{\partial }{\partial \lambda}F_{2,2}(x_1, x_2, v_1, v_2,l_{12}, l_{21})=
v_1\frac{\partial }{\partial x_1}F_{2,2}(x_1,x_2, v_1, v_2,l_{12},l_{21})
+v_2\frac{\partial }{\partial x_2}F_{2,2}(x_1,x_2,v_1, v_2,l_{12 },l_{21})\\
+2(\frac{{{|l}_{12}|}^2}{{(x_1-x_2)}^3}+\frac{{{|l}_{21}|}^2}{{(x_2-x_1)}^3})\frac{\partial }{\partial v_2}F_{2,2}(x_1,x_2,v_1, v_2,l_{12}, l_{21})\\
+2\int{dx_3dv_3{dl}_{13}dl_{31}dl_{23}dl_{32}}(\frac{{{|l}_{13}|}^2}{{(x_1-x_3)}^3}+\frac{{{|l}_{31}|}^2}{{(x_3-x_1)}^3})\frac{\partial }{\partial v_1}F_{3,6}(x_1, x_2,x_3,v_1,v_2,v_3,l_{12},l_{21},l_{13},l_{31},l_{23},l_{32})\\
+2\int{dx_3dv_3{dl}_{13}dl_{31}dl_{23}dl_{32}}(\frac{{{|l}_{23}|}^2}{{(x_2-x_3)}^3}+\frac{{{|l}_{32}|}^2}{{(x_3-x_2)}^3})\frac{\partial }{\partial v_2}F_{3,6}(x_1, x_2,x_3,v_1, v_2,v_3,l_{12},l_{21},l_{13},l_{31},l_{23},l_{32}).
\end{gathered}
\end{equation}
\end{widetext}

The above construction demonstrates the relationship between the distribution functions of the $s$-particles to $(s+1)$ interacting particles. 
These equations concern parametric evolution in time, such that application of the BBGKY hierarchy to the Pechukas model extends the hierarchy to parametric driven evolution of a quantum system, well suited for adiabatic systems. The coupled differential equations determine the kinetics of the distribution functions associated to the level dynamics of the Pechukas gas. The next step is to make the approximation that the distribution functions of the system can be expressed as a product of $F_{1,0}\left(x_1,v_1\right)$ distributions.

Taking into account the chain for $s=1$ we introduce a factorisation approximation based on the independence of the the set of coordinates ${x_m, v_m}$ and the set of relative angular momenta terms ${l_{mn}}$ such that we can construct the probability distribution functions of $x_m$ and $v_m$ that are independent of the probability distribution functions of $l_{mn}$. As a consequence, the reduced distribution $F_{s,s(s-1)}$ can be factorised in terms of the one-particle distribution and the distribution of $l_{mn}$ separately. Under the approximation,  the two-particle reduced distribution function describing the average density of the levels of a two particle system, $F_{2,2}\left(x_1,x_2,v_1,\ v_2,l_{12}, l_{21}\right)$ can be factorised in terms of the one particle distributions,  $F_{1,0}\left(x_1,v_1\right)$, $F_{1,0}\left(x_2,v_2\right)$ and the reduced distribution function concerning only the relative angular momentum terms $h(l_{12,\ \ }l_{21})$, (as defined in Appendix A)  with negligible contributions from the mixed terms. This is expressed below:

\begin{equation}
\begin{gathered}
F_{2,2}(x_1,x_2,v_1,v_2,l_{12},l_{21})\approx \\
F_{1,0}(x_1, v_1)F_{1,0}(x_2, v_2)h(l_{12},l_{21}).
\end{gathered}
\end{equation}

Under this approximation, Eq. (10) can be transformed in a way that precisely reflects an effective mean field theory, where the definitions of $F_{1,0}\left(x_1,v_1\right)$, $F_{1,0}\left(x_2,v_2\right)$ and $h(l_{12,\ \ }l_{21})$ are expressed in the same way as the generalised $F_{s,s\left(s-1\right)}$ as defined in Eq. (9).




Substituting Eq. (12) into Eq. (10) and using the product rule under the integral and taking $\left(\frac{\partial }{\partial v_1}F_{1,0}\left(x_1,\ v_1\right)\right)$ out from under the integral we obtain the following:
\begin{widetext}
\begin{equation}
\begin{gathered}
\frac{\partial }{\partial \lambda}F_{1,0}(x_1,v_1)=v_1\frac{\partial }{\partial x_1}F_{1,0}(x_1,v_1)\\
+2\frac{\partial }{\partial v_1}F_{1,0}(x_1, v_1)\int{dx_2dv_2dl_{12}dl_{21}}(\frac{{{|l}_{12}|}^2}{{(x_1-x_2)}^3}+\frac{{{|l}_{21}|}^2}{{(x_2-x_1)}^3})F_{1,0}(x_2, v_2)h(l_{12},l_{21}),
\end{gathered}
\end{equation}
\end{widetext}

by using the approximation we reduce the chain after breaking it at the first link such that only the $F_{1,0}\left(x_1,v_1\right)$, $F_{1,0}\left(x_2,v_2\right)$ and $h(l_{12,\ \ }l_{21})$ distributions are kept, hence simplifying to a one-body problem. 


\section{Accuracy of the Factorisation Approximation}
\noindent In this section, we estimate the accuracy of the factorisation approximation as this is important to know the validity of the effective mean field approximation. 

 Taking the definition for $F_{N,N\left(N-1\right)}\left(x_m,v_n,l_{mn}\right)$ as in Appendix A and evaluating the integral in Eq. (9), up to $s=2$ we obtain the distribution for $F_{2,2}\left(x_1,x_2,v_1,\ v_2,l_{12}, l_{21}\right)$ expressed as an average of the product of $\delta$ functions. Similarly, the product of the distributions $F_{1,0}\left(x_1,\ v_1\right),\ F_{1,0}(x_2,\ v_2)$ and $h(l_{12,\ \ }l_{21})$, determined from evaluating Eq. (9) with the definition of  $F_{N,N\left(N-1\right)}\left(x_m,v_n,l_{mn}\right)$ given in Appendix A, gives the product of $\delta$ functions averaged separately, with varying normalisation constants. 

Therefore, the error involved in verifying whether the factorisation approximation holds, is bounded by the normalisation constants used in Eq. (9). 
%
The normalisation constant concerning the reduced distribution for $l_{mn}$ takes the same form as in its counterpart in the two-particle reduced distribution. This is a consequence of the fact that the $l_{mn}$ term from $h(l_{12}, l_{21})$  comes from a system associated to the $F_{2,2}\left(x_1,x_2,v_1,\ v_2,l_{12}, l_{21}\right)$ distribution function. We determine the relative error, ${\cal{E}}_r := {\cal{E}}_r(x_1,x_2,v_1, v_2,l_{12}, l_{21})$ , such that it is bounded by the normalisation constants as given by the following expression:
\begin{widetext}
\begin{equation}
\begin{gathered}
{\cal{E}}_r :=\frac{{F_{1,0}(x_1, v_1)F_{1,0}(x_2, v_2)h(l_{12},l_{21})-F}_{2,2}(x_1,x_2,v_1, v_2,l_{12},l_{21})}{F_{2,2}(x_1,x_2,v_1, v_2,l_{12}, l_{21})},
\end{gathered}
\end{equation}
\end{widetext}

\noindent Substituting the expressions for the reduced distribution funcions, we obtain the following bound on the relative error:
\begin{equation}
\begin{gathered}
{\cal{E}}_r \leqslant \frac{N-(N-1)}{(N-1)}=O(\frac{1}{N}),
\end{gathered}
\end{equation}

For the limit that $N\to \infty $ we find that ${\cal{E}}_r$  decays asymptotically. 

We extend this further to consider the factorisation approximation for a general $s$-particle distribution function $F_{s,s(s-1)}$ such that it can be factorised as $s$ one particle distributions and $\frac{s\left(s-1\right)}{2}$  number of $h$ distributions. Using the same idea as the case for $F_{2,2}(x_1,x_2,v_1, v_2,l_{12},l_{21})$ we consider the way the normalisations constants will differ and the ${\cal{E}}_r$ between them. Taking the same approach, for the generalised factorisation for an $s$-particle distribution function composed as the product of $s$ one-particle distributions functions.



\noindent Then ${\cal{E}}_r$  is bounded by the following:
\begin{widetext}
\begin{equation}
\begin{gathered}
{\cal{E}}_r \leqslant \frac{N^s}{N(N-1)\dots (N-s-1)}.\frac{{((N^2-N)(N^2-N-1))}^{\frac{s(s-1)}{2}}}{(N^2-N)(N^2-N-1)\dots {(N}^2-N-s(s-1)-1)}-1=O(\frac{1}{N}),
\end{gathered}
\end{equation}
\end{widetext}

from these results it can be inferred that to solve the BBGKY chain for the Pechukas equations at the first link, only distributions functions for  $F_{1,0}\left(x_1,\ v_1\right)$, $F_{1,0}\left(x_2,\ v_2\right)$ and $h(l_{12,\ \ }l_{21})$ are required. The same rationale can be extended for higher order interactions in writing these distribution functions as products of one-particle distribution functions. We further illustrate this model on a two-qubit system. 

\section{A Two Qubit System}
\noindent We consider a two qubit system described by the Ising model, in order to test the BBGKY hierarchy for the Pechukas equations. We take the TFIH as the Hamiltonian that governs the two qubit system:

\begin{equation}
\begin{gathered}
\mathcal{H}(\lambda (t))=J{\sigma }^z_1{\sigma }^z_2+\lambda {Zh}_1{\sigma }^x_1+\lambda {Zh}_2{\sigma }^x_2,\\
\end{gathered}
\end{equation}

\noindent for the case that $J>0$ the interaction favors antiferromagnetism whereas for $J<0$ it favors ferromagnetism, we take random values for $J$, Gaussian distributied with mean 0 and standard deviation 1, reflecting the different initial conditions.  When $J\gg \lambda {Zh}_1,\lambda {Zh}_2$ the system is in the ground state. The perturbation matrix defined by $\lambda ZH_{b}=\lambda {Zh}_1{\sigma }^x_1+\lambda {Zh}_2{\sigma }^x_2$. From this we obtain the values for $x_n$ from the eigenvalues of the system given by $x_n\left(\lambda \right)=E_n\left(\lambda \right)=\langle n\left|{\mathcal{H}}\right|n\rangle$ the variables for velocity is determined by the following, $v_n(\lambda )=\langle n\left|Z{\mathcal{H}}_b\right|n\rangle$ and relative angular momentum, $l_{mn}$ using its definition that, $l_{mn}\left(\lambda \right)=\left(E_m\left(\lambda \right)-E_n\left(\lambda \right)\right)\langle m\left|Z{\mathcal{H}}_b\right|n\rangle$. ${\sigma }^z_j$ and ${\sigma }^x_j$ represent the corresponding Pauli matrices for the $j$-th qubit.
%

The Hamiltonian reads in matrix form:
\begin{widetext}
\begin{equation}
\begin{gathered}
\mathcal{H}\left(\lambda \left(t\right)\right)=J\left( \begin{array}{cccc}
1 & 0 & 0 & 0 \\ 
0 & -1 & 0 & 0 \\ 
0 & 0 & -1 & 0 \\ 
0 & 0 & 0 & 1 \end{array}
\right)+\lambda Zh_1\left( \begin{array}{cccc}
0 & 0 & 1 & 0 \\ 
0 & 0 & 0 & 1 \\ 
1 & 0 & 0 & 0 \\ 
0 & 1 & 0 & 0 \end{array}
\right)+\lambda Zh_2\left( \begin{array}{cccc}
0 & 1 & 0 & 0 \\ 
1 & 0 & 0 & 0 \\ 
0 & 0 & 0 & 1 \\ 
0 & 0 & 1 & 0 \end{array}
\right).
\end{gathered}
\end{equation}
\end{widetext}

Diagonalising the Hamiltonian and using their respective definitions, we determine the values for $x_n,\ v_n,\ l_{mn}$, necessary to construct the distribution functions for $f_1(\xi,\omega)$, $f_1(\xi',\omega')$, $f_2(\xi, \xi', \omega, \omega',l ,l')$ and $h(l,\ l')$. Where $\xi, \xi', \omega, \omega',l ,l'$ are the running variables of the probability distribution functions parameterising the coordinates $x_n,\ v_n,\ l_{mn}$ respectively.



%
%
The coordinates for $x_n$ are of the form $J+\lambda H_n$. Given that the values for $J$ are Gaussian distributed denoted by, $J\sim \mathcal{N}\left(\mu ,\sigma \right)$ the values for each $x_n$ are also Gaussian distributed varying only by a translation by $H_n$ hence,  $x_n\sim \mathcal{N}\left(\mu +\lambda H_n, \sigma\right)=\mathcal{N}(\lambda H_n, 1):= {\tilde{\mathcal{N}}}_n$, with the same mean and standard deviation where $H_n=\{-h_1-h_2,\ {-h}_1+h_2,h_1-h_2,h_1+h_2\}$. The values for $v_n$ are deterministic, we define them as $v_n\sim {\delta }_{H_n}$. We observe that the terms describing $l_{mn}$ determined from its definition are described by the product of the translated Gaussian distributions with the Dirac distributions for $H_n$, which we denote  by $L_{i,j}$. Using these definitions we build the distributions for $f_1\left(\xi, \omega \right),\ f_1(\xi',\omega')$, $f_2(\xi,\xi',\omega,\omega', l, l')$ and $h(l,\ l')$ as given below:
\begin{equation}
\begin{gathered}
f_1(\xi, \omega)=\frac{1}{4}\sum^4_{n=1}{\tilde{\mathcal{N}}_n(\xi){\delta }_{H_n}}(\omega)\\
h(l,l')=\frac{10!}{12!}\sum_{i\neq j}{L_{ij}}(l)\sum_{i^{'}\neq i,j'}{L^{'}_{i',j'}}(l')\\
f_2(\xi,\xi',\omega,\omega',l,l')=\\
\frac{10!}{4!12!}\sum_{i\neq j}{L_{ij}(l)\sum_{i^{'}\neq i,j'}{L^{'}_{i',j'}}}(l')\\
\times \sum^4_{ \begin{array}{c}
n\neq n^{' }\\ 
n,n^{'}=1 \end{array}
}{\tilde{\mathcal{N}}_n(\xi){\delta }_{H_n}}(\omega)\tilde{\mathcal{N}}_{n^{'}}(\xi'){\delta }_{H_{n^{'}}}(\omega'),
\end{gathered}
\end{equation}

substituting the definitions in Eq. (14), we analytically determine ${\cal{E}}_r$ for this system. We bound 
${\cal{E}}_r$  from above and below to examine the applicability of the factorisation approximation. The bounds have been calculated in Appendix B. We obtain:

\begin{equation}
\begin{gathered}
\frac{1}{8}e^{2\lambda {\xi'}(h_1-h_2)}\le {\cal{E}}_r \le \frac{1}{2}+\frac{1}{2}e^{2\lambda {\xi'}(h_1+h_2)},
\end{gathered}
\end{equation}

using these bounds, we find that the factorisation approximation does not hold, this is expected because to approximate well, a probability distribution function as a product of one-particle distribution functions, the system must be essentially factorisable such that the level dynamics can be treated as independent of each other. In the specific case of the two qubit system the qubits are not independent as they strongly interact, via $J$ terms. In order to explore this error further we test the system numerically using cloud dynamics and determine the distribution functions associated to the interactions. We draw 100 trial of $J\sim \mathcal{N}(0,1)$ terms, diagonalising the system to determine its eigenvalues, as demonstrated in Fig. 1. Choosing a large bias such that $Z=10$, we take $h_1$ as 0.01 and $h_2$ as 0.02, keeping these values small so as to reduce their impact on ${\cal{E}}_r$. However, to explore the dynamics of these values we diagonalise the Hamiltonian for $h_1$ as 0.1 and $h_2$ as 0.2 such that the perturbation is of the same order of the values of $J$ associated to the unperturbed Hamiltonian as in Eq. (17). The results are shown in Fig. 1.

 \begin{figure*}


 \begin{center}
  \includegraphics[scale=0.60]{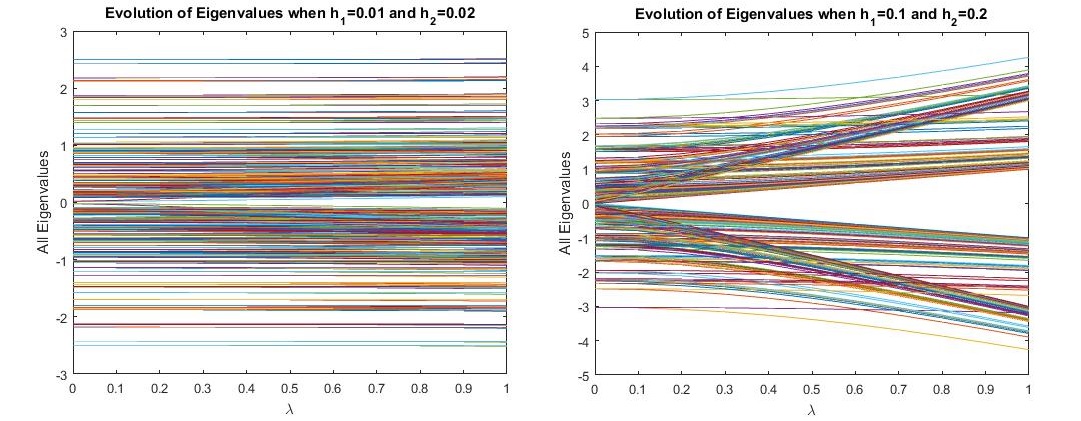}
\end{center}
  \caption{Evolution of eigenvalues: all the eigenvalues of Hamiltonian Eq. (18) for 100 simulations with random initial conditions obtained from the different values of $J$. These eigenvalues are of the form $J+\lambda H_n$, they are Gaussian distributed as $J$ is Gaussian distributed with mean 0 and standard deviation 1, through their evolution in $\lambda$ from 0 to 1 in steps of 0.1. When the perturbation is much weaker than the interaction $J$, the system stays close to its ground state. When the perturbation is of the same order as $J$, the eigenvalues deviate from an initially Gaussian distribution, evolving into four distinct peaks.}
\end{figure*}

%
%

Further to this, we construct the normalised distribution functions for $f_2(\xi,\xi',\omega,\omega', l, l')$ and that of  $f_1\left(\xi, \omega \right),\ f_1(\xi',\omega')$ and $h(l,\ l')$ in order to test the factorisation approximation for the first link of the BBGKY hierarchy using the Pechukas model. To build these distribution functions, we split the evolution of parameter, $\lambda$ interval in 0.1 from initial time at 0 and final time at 1, each of these distributions have been normalised. We compute $v_n$  by taking $x_{n+1}-x_n$ and dividing it through by the $\lambda$ interval step of 0.1. $l_{mn}$ is computed using its definition. As a result the distribution for $f_1\left(\xi, \omega \right)$ can be computed and the results shown in Fig. 2.

\begin{figure*}
  \begin{center}
    \includegraphics[scale=0.34]{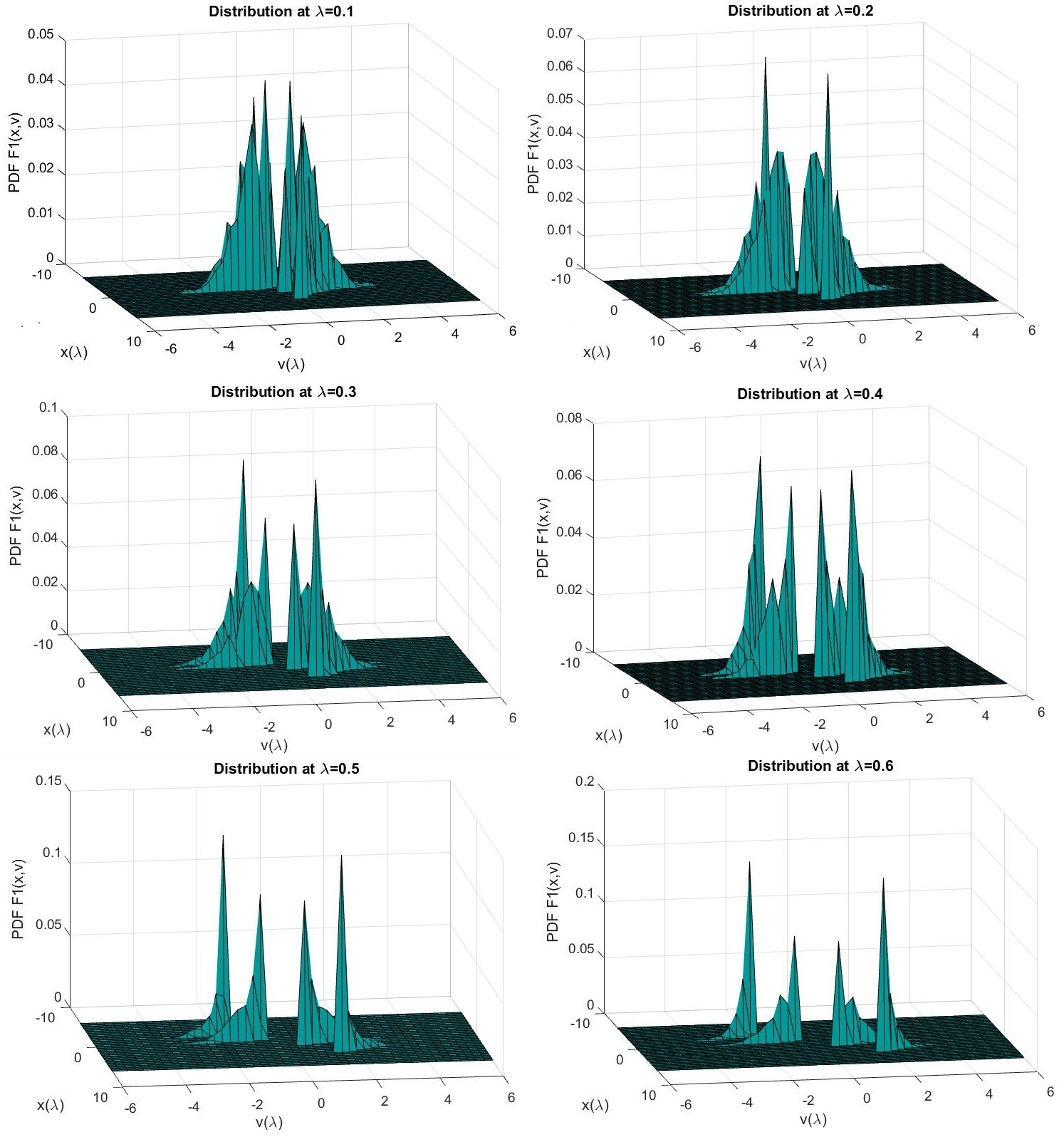}
    \includegraphics[scale=0.34]{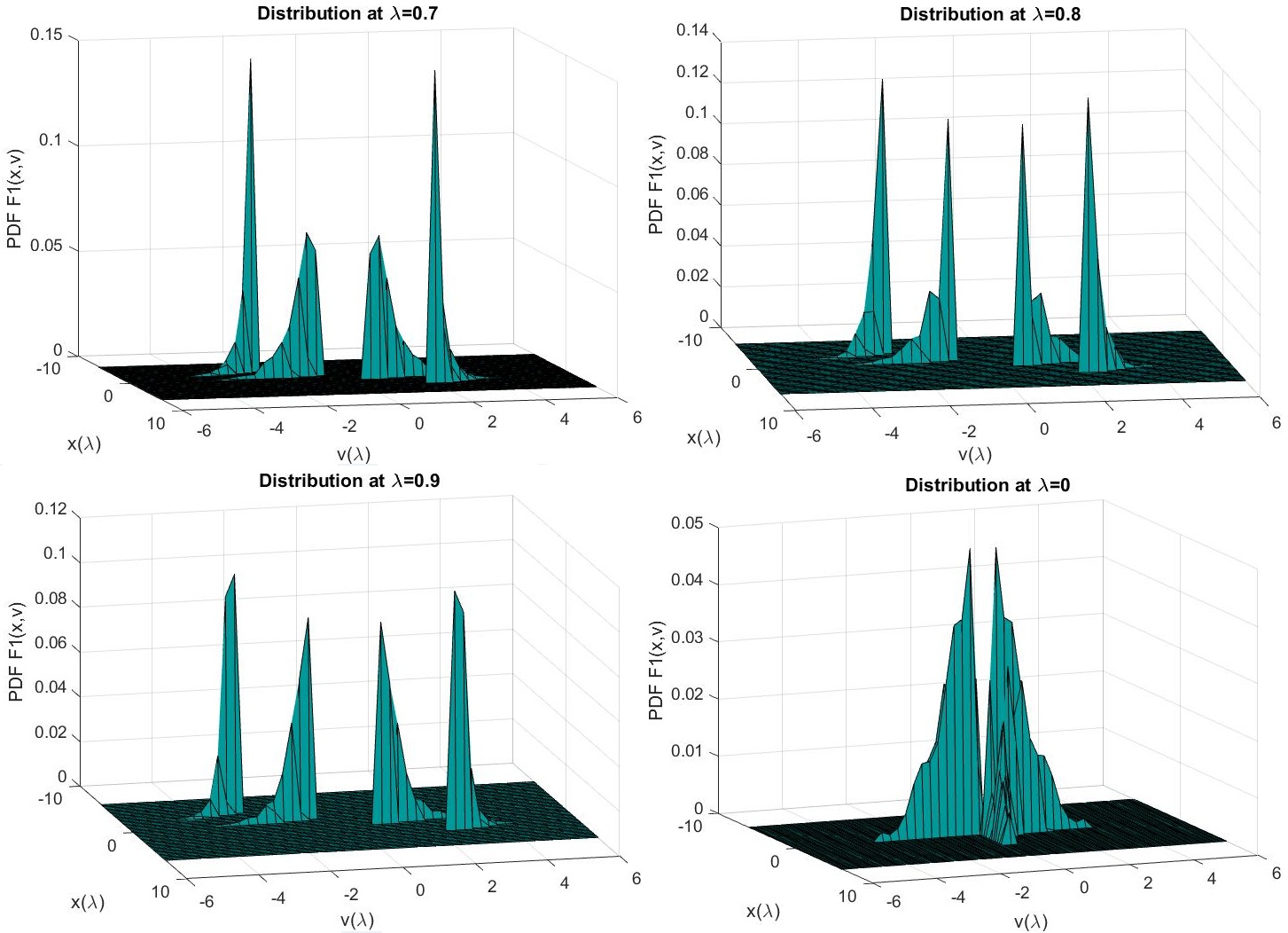}   
  \end{center}
  \caption{One-particle distributions: the evolution of $F_{1,0}\left(x_1,\ v_1\right)$ as  $\lambda$ increases. Initially Gaussian distributed about a single peak, as $\lambda$ increases, it settles into 4 equally distributed peaks due to the large perturbation. The velocities are deterministic so the distribution is centred around the four velocity points.}
 \label{fig:distribution1}
\end{figure*}

From the distribution functions we determine ${\cal{E}}_r$ using Eq. (14),  it is evident that the factorisation approximation does not hold for a two qubit system as discussed above. We use Eq. (14) to determine ${\cal{E}}_r$ and how it varies through the evolution of $\lambda$, considering nonzero points between the factorised distributions and that of  $f_2(\xi,\xi',\omega,\omega', l, l')$. The results are presented in the Table 1 below:

\begin{widetext}
\begin{tabular}{|p{0.5in}|p{0.5in}|p{0.5in}|p{0.5in}|p{0.5in}|p{0.5in}|p{0.5in}|p{0.5in}|p{0.5in}|p{0.5in}|} \hline 
$\lambda $ & 0.1 & 0.2 & 0.3 & 0.4 & 0.5 & 0.6 & 0.7 & 0.8 & 0.9 \\ \hline 
$\langle{\cal{E}}_r \rangle$ & 0.668 & 3.203 & 0.578 & 1.18 & 0.335 & 1.54 & 0.4803 & 0.0659 & 0.366 \\ \hline 
SD of ${\cal{E}}_r $  & 0.00127 & 0.00393 & 0.00119 & 0.00216 & 0.00120 & 0.00312 & 0.00142 & 0.00109 & 0.00153 \\ \hline 
\end{tabular}

{\noindent Table 1: the average ${\cal{E}}_r$  and its standard deviation (SD) are described through time up to 3 significant figures, to determine the accuracy of the factorisation approximation, using the Pechukas model for a two qubit system.}
\end{widetext}

We note that ${\cal{E}}_r$'s standard deviation remains below 0.005 throughout the evolution of the adiabatic parameter. We observe, anomalous averaged relative errors as in the cases for $\lambda$  being 0.2 ,0.4 and 0.6 that which do not fall in the range of the analytic bounds determined from Eq. (20). This is a result of the sample being taken from 100 trials. However  ${\cal{E}}_r$ follows the prediction of Eq. (15) with ${\cal{E}}_r$ expected to be 1 for a two qubit system as well as Eq. (20) which suggests an exponential growth in the upper bound for a two qubit system with a minimum of $\frac{1}{2}$. 
We predict that as $N$ grows the system settles close to the mean field behavior. Though the factorisation approximation has not been numerically tested for large $N$, we have shown that ${\cal{E}}_r$ scales as $1/N$, suggesting that it is possible to reduce the BBGKY chain to a factorisation of $F_{1,0}\left(x_1,\ v_1\right)$ distributions. We leave the numerical demonstration of this for future work as it is beyond the scope of the present mainly analytical study.

\section{Evolution of Occupation Numbers and the Density Matrix}

\noindent In this section we establish the relationship between the occupation numbers and the level dynamics through the Pechukas model. Details of this section are provided in Appendix C. As the wave function on a Hilbert space can be expressed as the sum of linear combination of eigenstates \cite{Requist2, Requist3} such that from the eigenstate coefficients $C_n(t)\in \mathbb{C}$ determines the occupation numbers as shown in Appendix C. 



\noindent The evolution of $C_n$ is given by the following: 

 \begin{equation}
\begin{gathered}
i\dot{C_m}(t)-C_m(t)E_m=\\
-i\sum_{n\neq m}{C_n(t)}\langle m(t)|\frac{\partial }{\partial \lambda }|n(t)\rangle \dot{\lambda(t)},
\end{gathered}
\end{equation}

\noindent where $E_m(t)$ are the eigenvalues of the system with quantum states $m(t)$ and $n(t)$. In order to evaluate the dynamics with respect to the level dynamics, it is necessary to determine $\sum_n{i}C_n(t)\langle m(t)|\frac{\partial }{\partial \lambda }|n\left(t\right)\rangle \dot{\lambda }$, where the term vanishes for $m(t)=n(t)$. 
We adopt the Pechukas model in order to express the evolution of $C_n$ in terms of variables describing level dynamics. The evolution of $\langle m(t)\left|\frac{\partial }{\partial \lambda }\right|n(t)\rangle$  reads:

 \begin{equation}
\begin{gathered}
\langle m(t)|\frac{\partial }{\partial \lambda}|n(t)\rangle=\frac{-l_{mn}}{{(x_m-x_n)}^2},
\end{gathered}
\end{equation}

\noindent substituting the result into Eq. (21), the dynamics of the occupation numbers are given by:
 \begin{equation}
\begin{gathered}
i\dot{C_m}(t)-C_m(t)x_m=i\dot\lambda(t) \sum_{n\neq m}{C_n}(t)\frac{l_{mn}}{{(x_m-x_n)}^2},
\end{gathered}
\end{equation}

The wave function can be entirely described as a function of time. 

For the simplified case that $\dot{\lambda }=0$, we solve this ordinary differential equation to find that $C_m\left(t\right)=C_m(0)e^{-i\int^t_0{x_m\left(s\right)ds}}$, describing a system where level crossings are not possible. For the case under consideration, Eq. (23) is inhomogeneous. If we define the following square matrices and vector:
\[X=\mathrm{diag}\left(x_1\dots x_n\right)\] 
$P=p_{mn}$ where $p_{mn}=\frac{l_{mn}}{{\left(x_m-x_n\right)}^2}$ and $p_{mm}=0$
\[C={\left(C_1(t)\dots C_n(t)\right)}^T\]

we can then obtain a set of coupled differential equation
 \begin{equation}
\begin{gathered}
i\frac{\partial }{\partial t}C=(X+i\dot{\lambda }P)C,
\end{gathered}
\end{equation}

in this case, $X$ is diagonal and $P$ is skew-Hermitian as  $l_{mn}=-l^*_{nm}$ thus diagonalisable. Let $M=(-iX+\dot{\lambda }P$). This matrix does not commute with itself for different time instances thus using the Peano-Baker series (PBS) described in Ref. \onlinecite{PBS} we find that the solution comes in the form:
 \begin{equation}
\begin{gathered}
\mathrm{C}(\mathrm{t}\mathrm{;}{\mathrm{t}}_0)\mathrm{=}\boldsymbol{\mathrm{1}}\boldsymbol{\mathrm{+}}\sum^{\infty }_{n=1}{{\mathcal{I}}_n(t)},
\end{gathered}
\end{equation}

\noindent where $t_0$ is the initial time and ${\mathcal{I}}_n$ is expressed as:
 \begin{equation}
\begin{gathered}
{\mathcal{I}}_{\mathrm{n}}(\mathrm{t}):=\\
\mathrm{\ }\int^t_{t_0}{M({\tau }_1)}\int^{{\tau }_1}_{t_0}{M({\tau }_2)}\dots \int^{{\tau }_{n-1}}_{t_0}{M({\tau }_n)d{\tau }_n\dots d{\tau }_1}.
\end{gathered}
\end{equation}

\noindent The above demonstrates that a solution exists for a general $\dot{\lambda}(t)$. For $\lambda$ being an adiabatic parameter we are interested in a nonzero constant $\dot{\lambda }$ such that the adiabatic parameter evolves slowly enough that the system is not excited from its eigenstate. In the case that the matrix $M(t)$ commutes with itself at each instant in time, we may use classic linear algebra to determine $C_n(t)$ at each instance through the relation:
 \begin{equation}
\begin{gathered}
C(t)={e}^{-i\int^t_0{M(s)ds}}C_0,
\end{gathered}
\end{equation}

This expression holds only when $M\left(t\right)$ can be approximated as constant, whereas using PBS an explicit solution can be evaluated for any $\dot{\lambda }\left(t\right)$, in order to determine $C_n(t)$ at each instant. 
We use the relationships established between occupation numbers and level dynamics in order to determine the density matrix through the Pechukas equations. Solving $C_n(t)$, we obtain the density matrix with entries of the form:
%
%
 \begin{equation}
\begin{gathered}
{\rho }_{nm}(t)=C^{*}_m(t)C_n(t),
\end{gathered}
\end{equation}

this can be determined by solving for $C\left(t\right)$ as in Eq.(23). In the case that the system collapses to that of an approximated constant matrix describing $M(t)$, we can substitute the solution described in (28) to obtain: 
 \begin{equation}
\begin{gathered}
{\rho }_{nm}(t)=\\
\sum^N_{k^{'},k=1}{C_{k^{'}}(t)C^{*}_k(t)A_{mk^{'}}{(t)A}^{*}_{nk}(t)e^{i\int^t_0{d^{*}_k(s)-d_{k^{'}}(s)ds}}},
\end{gathered}
\end{equation}

then it can be simply expressed as a tensor product in order to describe the evolution of the density matrix as in Eq. (30). 
 \begin{equation}
\begin{gathered}
\rho (t)=C(t)\otimes C^{*}(t),
\end{gathered}
\end{equation}

as a result, we obtain the final relationship between the density matrix evolution in time and the dynamics of the occupation numbers. These are expressed in terms of the Pechukas model and so the entire system can be described through the evolution of the levels. 

\section{Discussion-Conclusions.}
\noindent We used the mapping to Pechukas model to determine the kinetic equations of motion given by the BBGKY hierarchy, in order to explore the level dynamics of a system. This procedure provides a fundamental extension of previous study which established the use of Pechukas model and the random matrix theory in equilibrium statistical mechanics of the gas.The application of the BBGKY hierarchy to the Pechukas model, extends the kinetic equations concerning the level dynamics to parametrically driven evolution of a quantum system which is especially convenient for the investigation of adiabatically evolving systems however the formalism is not strictly adiabatic and is applicable to a general system of parametric time evolution with arbitrary time dependence in $\lambda$.

We also explored the factorisation approximation that the $s$-particle reduced probability distribution functions can be constructed from a product of $s$ one-particle distributions. The insight to study this is given by the fact that the coordinates in the Pechukas equations are independent and so may lead to effectively independent probability distribution functions, reducing the many body systems to that of a one-body system. This is a great simplification as it amounts to solving the BBGKY hierarchy by solving just the one-body system. All the information of the level dynamics can be determined from the one-particle distribution functions. To test the factorisation approximation we have analytically considered the way the factorisations vary from the many particle probability distribution functions giving an effective mean field theory approximation. We find that the relative error ${\cal{E}}_r$  decays asymptotically as $O\left(\frac{1}{N}\right)$ as the number of the interacting particles tend to infinity. This gives confidence that for systems with large number of particle interactions, the approximation holds. 

To illustrate the theory we considered the simplest possible system of two qubits, and compared it with the exact solution of the Hamiltonian. Using the eigenvalues in accordance with the Pechukas equations the velocities and relative angular momenta can be determined. From these, the distribution functions involved in the first chain of the BBGKY hierarchy were constructed in order to test the factorisation approximation. For the case of two qubits, the factorisation approximation is not accurate and so it is found the probability distribution functions do not factorise such that the energy levels for any given value of the parameter $\lambda$ are not mutually independent. A comprehensive comparison of larger systems is impractical as it is harder to diagonalise the corresponding Hamiltonians and it is beyond the scope of the present study.

We determined the relation between the level dynamics and the evolution of occupation numbers, where time appears explicitly for the first time. From these occupation numbers we are able to describe the entire wavefunction using the Pechukas equations. In order to determine the occupation numbers, we must solve the corresponding ordinary differential equations. Here we have found that the matrix describing the coupled differential equations does not commute with itself at different time instances and so it is necessary to consider alternative methods to solve for the occupation numbers. 
Using the developed theory, it can be possible to determine the ground state of large systems efficiently and hence to demonstrate that the adiabatic quantum computing is an alternative to quantum computing. 

\section*{Acknowledgments} We are grateful to thank Peter Mason, Sergey Savel'ev and Alexander Veselov for the valuable discussions that greatly improved the manuscript.  This work has been supported by EPSRC through the grant EP/M006581/1.

\section*{Appendix A: Derivation of the BBGKY Hierarchy for the Pechukas Model.}
The BBGKY chain for the Pechukas equations govern the level dynamics of the system. In turn this determines the way the particles interact through considering the associated reduced distribution functions relating that of $s$-particles to $(s+1)$ particles. The derivation starts with defining the distribution function:
\noindent Consider the distribution with dynamic variables $x_m,v_n$ and $l_{mn}$:  averaging over $\xi ,\ \omega ,\ \Lambda $

\begin{equation}
\begin{gathered}
F_{N,N(N-1)}(x_1, \dots,x_n, v_1, \dots,v_n,l_{12}, \dots, l_{mn})\\
=\langle \prod_m{\delta (x_m-{\xi }_m)\delta (v_m-{\omega }_m)}\prod_{m \neq n}{\delta (l_{mn}-{\Lambda}_{mn})}\rangle.
\end{gathered}
\end{equation}

\noindent The averaging procedure is described through:
\begin{equation}
\begin{gathered}
\langle F_N(x^{\lambda},v^{\lambda},l^{\lambda},\lambda;x^0,\ v^0,\ l^0)\rangle :=\\
\frac{1}{|S|} \sum_{x^0,\ v^0,l^0\in S} 〖f (x^{\lambda},v^{\lambda},l^{\lambda},\lambda;x^0,\ v^0,\ l^0)〗,
\end{gathered}
\end{equation}

\noindent Where $|S|$ is the size of $S$, being the set of all initial conditions $(x^0,\ v^0,l^0)$ and $F_N(x^{\lambda},v^{\lambda},l^{\lambda},\lambda;x^0,\ v^0,\ l^0)$ denotes the function evaluated at $(x^{\lambda},v^{\lambda},l^{\lambda},\lambda)$, the propagated coordinates obtained through the Pechkuas equation up to evolution of $\lambda$. We take $F_{N,N(N-1)}:= F_{N,N(N-1)}(x_1, \dots,x_n, v_1, \dots,v_n,l_{12}, \dots, l_{mn})$ All distribution functions are symmetric with respect to permutations of arguments \cite{Pechukas, Yukawa1, Haake, Balescu}.

\noindent Here $x_m,v_n$ and $l_{mn}$ are independent coordinates which describe the centre frame and $\xi ,\ \omega $ and $\mathrm{\Lambda }$ are shifted coordinates from this centre frame. Taking a total derivative of this distribution with respect to the adiabatic parameter $\lambda$, we obtain the following:
\begin{widetext}
\begin{eqnarray}
\nonumber
\frac{dF}{d\lambda}
=\sum_{m\neq m^{'}}{\langle \prod_{m^{'}}{\frac{\partial }{\partial {\xi }_m}\delta (x_{m^{'}}-{\xi }_{m^{'}})\dot{{\xi }_m}\delta (v_{m^{'}}-{\omega }_{m^{'}})}}
\prod_{m^{'}n^{'}}{\delta (l_{m^{'}n^{'}}-{\mathrm{\Lambda }}_{m^{'}n^{'}})}\rangle\\
\nonumber
+\sum_{m\neq m^{'}}{\langle \prod_m{\delta (x_{m^{'}}-{\xi }_{m^{'}})}\frac{\partial }{\partial {\omega }_m}\delta (v_{m^{'}}-{\omega }_{m^{'}})}\dot{{\omega }_m}\prod_{m^{'}n^{'}}
{\delta (l_{m^{'}n^{'}}-{\mathrm{\Lambda }}_{m^{'}n^{'}})}\rangle\\
+\langle \prod_{m^{'}}{\delta (x_{m^{'}}-{\xi }_{m^{'}})\delta (v_{m^{'}}-{\omega }_{m^{'}})}\sum_{m\neq m^{'}, n\neq n^{'}}{\prod_{m^{'}n^{'}}}{\frac{\partial }{\partial {\mathrm{\Lambda }}_{mn}}\delta (l_{m^{'}n^{'}}-{\mathrm{\Lambda }}_{m^{'}n^{'}})}\dot{{\mathrm{\Lambda }}_{mn}}\rangle
+\frac{\partial F}{\partial \lambda}.
\end{eqnarray}
\end{widetext}
\noindent Given that $\frac{\partial }{\partial \xi }=-\frac{\partial }{\partial x}$ , $\frac{\partial }{\partial \omega }=-\frac{\partial }{\partial v}$, $\frac{\partial }{\partial \mathrm{\Lambda }}=-\frac{\partial }{\partial l}$ where '.' describing differentiation with respect to $\lambda $ as described in (1). We substitute this into the total derivative with the related Pechukas equations through the chain rule with respect to time to obtain:
\begin{widetext}
\begin{eqnarray}
\nonumber
\frac{dF}{d\lambda}
=-\sum_{m}{v_{m}\frac{\partial }{\partial {\xi }_m}F_{N,N(N-1)}}
-\sum_m{\frac{\partial }{\partial v_m}2\sum_{m\neq n}}\frac{{{|l}_{mn}|}^2}{{(x_m-x_n)}^3}F_{N,N(N-1)}\\
-\sum_{mn}{\frac{\partial }{\partial l_{mn}}}\sum_{k\neq m,n}{l_{mk}l_{kn}}(\frac{1}{{(x_m-x_k)}^2}-\frac{1}{({x_k-x_n)}^2})F_{N,N(N-1)}
+\frac{\partial }{\partial \lambda}F_{N,N(N-1)}.
\end{eqnarray}
\end{widetext}

\noindent Applying Liouville's theorem that the number of particles at the start stays constant in the system as time evolves, preserving probability along the trajectory in phase space\cite{Pechukas, Yukawa1, Haake, Balescu}:

\begin{equation}
\begin{gathered}
\frac{dF}{d\lambda}=0.
\end{gathered}
\end{equation}

\noindent From this we rearrange the total derivative to express $\frac{\partial }{\partial \lambda}F_{N,N\left(N-1\right)}\left(x_m,v_m,l_{mn}\right)$ by the following:
\begin{widetext}
\begin{eqnarray}
\nonumber
\frac{\partial }{\partial \lambda}F_{N,N(N-1)}=
\sum_m{v_m\frac{\partial }{\partial x_m}F_{N,N(N-1)}}
+\sum_m{\frac{\partial }{\partial v_m}}2\sum_{m\neq n}{\frac{{{|l}_{mn}|}^2}{{(x_m-x_n)}^3}}F_{N,N(N-1)}\\
+\sum_{mn}{\frac{\partial }{\partial l_{mn}}}\sum_{k\neq m,n}{l_{mk}l_{kn}}(\frac{1}{{(x_m-x_k)}^2}-\frac{1}{({x_k-x_n)}^2})F_{N,N(N-1)}.
\end{eqnarray}
\end{widetext}

\noindent In the scheme of BBGKY, we consider $s$ number of particles where $s\le \{1,\dots N\}$ in order to build up the chain. For this we consider the way each term of the distribution is affected. The $s$-particle distribution function is thus given by:
\begin{widetext}
\begin{equation}
\begin{gathered}
F_{s,s\left(s-1\right)}{{:=}}
\frac{N!}{(N-s)!}.\frac{(N^2-N)!}{(N^2-N-s(s-1)!}
\int{dx_{s+1}\dots dx_ndv_{s+1}\dots dv_ndl_{s+1,s}\dots dl_{n,n(n-1)}F_{N,N(N-1)}}.
\end{gathered}
\end{equation}
\end{widetext}

\noindent The normalisation constants in the front of the integral comes from the combinatorics of $x_m$ and $v_m$ for $\frac{N!}{(N-s)!}$ With $N!$ counts the total number of combinations in both $x_m$ and $v_m$ and $\left(N-s\right)!$ counts the number of combinations of particles not included in the distribution. Similarly, for $l_{mn}$ we have the total number of possible values in $l_{mn}$ determined by $\left(N^2-N\right)!$. When dictated by $s$ there are $s(s-1)$ possible values in the distribution giving rise to the normalisation constant $\frac{\left(N^2-N\right)!}{\left(N^2-N-s(s-1)\right)!}$, included in the definition for $F_{s,s\left(s-1\right)}$. These come from the symmetry in the distribution functions with respect to permutations in their arguments. 

\noindent Considering the $s$-particle distribution in the above relation for Eq. (36) we obtain:
\begin{widetext}
\begin{eqnarray}
\nonumber
\frac{\partial }{\partial \lambda}F_{s,s(s-1)}=\frac{N!}{(N-s)!}.\frac{(N^2-N)!}{(N^2-N-s(s-1))!}\int{dx_{s+1}\dots dx_ndv_{s+1}\dots dv_ndl_{s+1,s}\dots dl_{n,n(n-1)}}\\
\nonumber
{\sum_m{v_m\frac{\partial }{\partial x_m}F_{N,N\left(N-1\right)}}}
+\sum_m{\frac{\partial }{\partial v_m}2\sum_{m\neq n}}{\frac{{{|l}_{mn}|}^2}{{(x_m-x_n)}^3}}F_{N,N(N-1)}\\
+\sum_{mn}{\frac{\partial }{\partial l_{mn}}\sum_{k\neq m,n}}{l_{mk}l_{kn}}(\frac{1}{{(x_m-x_k)}^2}-\frac{1}{({x_k-x_n)}^2})F_{N,N(N-1)}\}.
\end{eqnarray}
\end{widetext}
\noindent Determining the way the first term is affected by the reduced distribution function concerning up to $s$-particle interactions is expressed by:
\begin{widetext}
\begin{eqnarray}
\nonumber
\frac{N!}{(N-s)!}.\frac{(N^2-N)!}{(N^2-N-s(s-1))!}\int{dx_{s+1}\dots dx_ndv_{s+1}\dots dv_ndl_{s+1,s}\dots dl_{n,n(n-1)}}\\
{\sum^s_{m=1}}{v_m\frac{\partial }{\partial x_m}F_{N,N(N-1)}}
+\int{dx_{s+1}\dots dx_ndv_{s+1}\dots dv_ndl_{s+1,s}\dots dl_{n,n(n-1)}}{\sum^N_{m=s+1}{v_m\frac{\partial }{\partial x_m}F_{N,N(N-1)}}}.
\end{eqnarray}
\end{widetext}


Using the Green's theorem the last term vanishes as the system tends to infinity, which reduces the expression to:
\begin{equation}
\begin{gathered}
\frac{N!}{(N-s)!}.\frac{(N^2-N)!}{(N^2-N-s(s-1))!}\\
\int{dx_{s+1}\dots dx_ndv_{s+1}\dots dv_ndl_{s+1,s}\dots dl_{n,n(n-1)}}\\
{\sum^s_{m=1}{v_m\frac{\partial }{\partial x_m}F_{N,N(N-1)}}}.
\end{gathered}
\end{equation}

\noindent Following this procedure, we determine the way $F_{s,s\left(s-1\right)}$ affects the second term in the relation for $\frac{\partial }{\partial \lambda}F_{N,N\left(N-1\right)}$ such that it concerns only the $s$-particle distribution and $(s+1)$ particle distribution as determined below:

\begin{widetext}
\begin{eqnarray}
\nonumber
\frac{N!}{(N-s)!}.\frac{(N^2-N)!}{(N^2-N-s(s-1))!}
\nonumber
\int{dx_{s+1}\dots dx_ndv_{s+1}\dots dv_ndl_{s+1,s}\dots dl_{n,n(n-1)}}\\
\nonumber
2{\sum^s_{m=1}{\sum^{m-1}_{n=1}{(\frac{{{|l}_{mn}|}^2}{{(x_m-x_n)}^3}+\frac{{{|l}_{nm}|}^2}{{(x_n-x_m)}^3})\frac{\partial }{\partial v_m}F_{N,N(N-1)}}}}\\
+2\int{dx_{s+1}\dots dx_ndv_{s+1}\dots dv_ndl_{s+1,s}\dots dl_{n,n(n-1)}}{\sum^s_{m=1}{(\frac{{{|l}_{m(s+1)}|}^2}{{(x_m-x_{s+1})}^3}+\frac{{{|l}_{(s+1)m}|}^2}{{(x_{s+1}-x_m)}^3})\frac{\partial }{\partial v_m}F_{N,N(N-1)}}}.
\end{eqnarray}
\end{widetext}

\noindent Finally, we determine the way taking $F_{s,s\left(s-1\right)}$ affects the last term in the relation for $\frac{\partial }{\partial \lambda}F_{N,N\left(N-1\right)}$, where we obtain the following:


\begin{widetext}
\begin{equation}
\begin{gathered}
\frac{N!}{(N-s)!}.\frac{(N^2-N)!}{(N^2-N-s(s-1))!}\\
\int{dx_{s+1}\dots dx_ndv_{s+1}\dots dv_ndl_{s+1,s}\dots dl_{n,n(n-1)}}
{\sum^s_{m=1}{\sum^{m-1}_{k=1}{\sum^{k-1}_{n=1}{l_{mk}l_{kn}(\frac{1}{{(x_m-x_k)}^2}-\frac{1}{({x_k-x_n)}^2})\frac{\partial }{\partial l_{mn}}}F_{N,N(N-1)}}}}\\+\int{dx_{s+1}\dots dx_ndv_{s+1}\dots dv_ndl_{s+1,s}\dots dl_{n,n(n-1)}}{\sum^s_{k=1}{\sum^{k-1}_{n=1}{l_{s+1k}l_{kn}(\frac{1}{{(x_{s+1}-x_k)}^2}-\frac{1}{({x_k-x_n)}^2})\frac{\partial }{\partial l_{(s+1)n}}}F_{N,N(N-1)}}}.
\end{gathered}
\end{equation}
\end{widetext}

\noindent Combining these expressions and simplifying with the definition for $F_{s,s\left(s-1\right)}$ we derive the BBGKY chain for the Pechukas equations as given by the following equations, where we denote $\boldsymbol{D}{\boldsymbol{l}}_{s+1}=\prod^s_{i=1}{dl_{s+1,i}dl_{i,s+1}}$:

\begin{widetext}
\begin{eqnarray}
\nonumber
\frac{\partial }{\partial \lambda}F_{s,s(s-1)}
=\sum^s_{m=1}{v_m\frac{\partial }{\partial x_m}F_{s,s(s-1)}}
+2\sum^s_{m=1}{\sum^{m-1}_{n=1}{(\frac{{{|l}_{mn}|}^2}{{(x_m-x_n)}^3}+\frac{{{|l}_{nm}|}^2}{{(x_n-x_m)}^3})\frac{\partial }{\partial v_m}F_{s,s(s-1)}}}\\
\nonumber
+2\int{dx_{s+1}dv_{s+1}\boldsymbol{D}{\boldsymbol{l}}_{s+1}}
\sum^s_{m=1}{(\frac{{{|l}_{m(s+1)}|}^2}{{(x_m-x_{s+1})}^3}+\frac{{{|l}_{(s+1)m}|}^2}{{(x_{s+1}-x_m)}^3})\frac{\partial }{\partial v_m}F_{s+1,s\left(s+1\right)}}\\
\nonumber
+\sum^s_{m=1}{\sum^{m-1}_{k=1}{\sum^{k-1}_{n=1}{l_{mk}l_{kn}(\frac{1}{{(x_m-x_k)}^2}-\frac{1}{({x_k-x_n)}^2})\frac{\partial }{\partial l_{mn}}}F_{s,s(s-1)}}}\\+\int{dx_{s+1}dv_{s+1}\boldsymbol{D}{\boldsymbol{l}}_{s+1}}\sum^s_{k=1}{\sum^{k-1}_{n=1}{l_{s+1k}l_{kn}(\frac{1}{{(x_{s+1}-x_k)}^2}-\frac{1}{({x_k-x_n)}^2})\frac{\partial }{\partial l_{(s+1)n}}}F_{s+1,s(s+1)}}.
\end{eqnarray}
\end{widetext}

\noindent This gives us the BBGKY hierarchy for the Pechukas model with respect to a full distribution concerning position, velocity and relative angular momentum. 

In constructing the factorisation approximation we define the reduced distribution function concerning the relative angular momentum terms for the level dynamics of $l_{1,2},l_{2,1}$ by $h(l_{1,2},l_{2,1})$. The distribution function is given by $h_{N,N(N-1)}(l_{1,2}\dots l_{n,m}\dots l_{m,n})=\langle\prod_{n \neq m}{\delta (l_{mn}-{\Lambda}_{mn})}\rangle$, analogously to that of $F_{N,N(N-1)}$. This takes a reduced distribution function up to the s-particle defined by the following:

\begin{widetext}
\begin{equation}
\begin{gathered}
 h_{s,s(s-1)}(l_{1,2},...,l_{s,s(s+1)},...,l_{s(s+1),s})
{{:=}}\frac{(N^2-N)!}{(N^2-N-s(s-1)!}\int{dl_{s+1,s}\dots dl_{n,n(n-1)}h_{N,N(N-1)}}.
\end{gathered}
\end{equation}
\end{widetext}

\noindent Given that this approximation only concerns the dynamics of $l_{1,2}, l_{2,1}$ we denote $h := h_{2,2}$ using the above definitions. 

\section*{Appendix B: Determining the relative error ${\cal{E}}_r$ of the two-particle distribution and it's extension to the two qubit system.}

We have tested the factorisation approximation for a simple case of two qubits.

\noindent Substituting the definitions given in Eq. (19) into Eq. (14) we analytically determine ${\cal{E}}_r$  for this system. We reduce the ${\cal{E}}_r$  as given in the expression below. Keeping concise, we omit the arguments of the distributions:

\begin{equation}
\begin{gathered}
{\cal{E}}_r=\frac{f_1(\xi, \omega).f_1(\xi', \omega').h(l,l')-f_2(\xi,\xi',\omega,\omega',l,l')}{f_2(\xi,\xi',\omega,\omega',l,l')},
\end{gathered}
\end{equation}

\noindent where these expressions are given as defined in Eq. (19)


\noindent Substituting these expressions, we reduce  ${\cal{E}}_r$ to the following:

\begin{widetext}
\begin{equation}
\begin{gathered}
{\cal{E}}_r=\frac{3}{2}\frac{\sum^4_{n=1}{\tilde{\mathcal{N}}_n(\xi){\delta }_{H_n}}(\omega) \sum^4_{n=1}{\tilde{\mathcal{N}}_n'(\xi'){\delta }_{H_n'}}(\omega')-\sum^4_{ \begin{array}{c}
n\neq n^{'} \\ 
n,n^{'}=1 \end{array}
}{\tilde{\mathcal{N}}_n(\xi){\delta }_{H_n}}(\omega)\tilde{\mathcal{N}}_{n^{'}}(\xi'){\delta }_{H_{n^{'}}}(\omega')}{\sum^4_{ \begin{array}{c}
n\neq n^{'} \\ 
n,n^{'}=1 \end{array}
}{\tilde{\mathcal{N}}_n(\xi){\delta }_{H_n}}(\omega)\tilde{\mathcal{N}}_{n^{'}}(\xi'){\delta }_{H_{n^{'}}}(\omega')}.
\end{gathered}
\end{equation}
\end{widetext}

\noindent Dropping the arguments such that $\tilde{\mathcal{N}}_i(\xi){\delta }_{H_i}(\omega):=\tilde{\mathcal{N}}_i{\delta }_{H_i}$ and similarly $\tilde{\mathcal{N}}_i(\xi'){\delta }_{H_i}(\omega'):=\tilde{\mathcal{N}}_i'{\delta }_{H_i'}$, this expression is simplified to the following by expanding these sums and cancelling common terms. 

\begin{widetext}
\begin{equation}
\begin{gathered}
{\cal{E}}_r=\frac{3}{2}\frac{\tilde{\mathcal{N}}_1{\delta }_{H_1}\tilde{\mathcal{N}}_{1^{'}}{\delta }_{H_{1^{'}}}+\tilde{\mathcal{N}}_2{\delta }_{H_2}\tilde{\mathcal{N}}_{2^{'}}{\delta }_{H_{2^{'}}}+\tilde{\mathcal{N}}_3{\delta }_{H_3}\tilde{\mathcal{N}}_{3^{'}}{\delta }_{H_{3^{'}}}+\tilde{\mathcal{N}}_4{\delta }_{H_4}\tilde{\mathcal{N}}_{4^{'}}{\delta }_{H_{4^{'}}}}{\sum^4_{ \begin{array}{c}
n\neq n^{'} \\ 
n,n^{'}=1 \end{array}
}{\tilde{\mathcal{N}}_n{\delta }_{H_n}}\tilde{\mathcal{N}}_{n^{'}}{\delta }_{H_{n^{'}}}}
+\frac{1}{2}.
\end{gathered}
\end{equation}
\end{widetext}

\noindent We bound the error from above through maximising the numerator with $4(\tilde{\mathcal{N}}_4(\xi){\delta }_{H_4}(\omega)\tilde{\mathcal{N}}_{4^{'}}(\xi'){\delta }_{H_{4^{'}}}(\omega'))$ where $(\tilde{\mathcal{N}}_4(\xi){\delta }_{H_4}(\omega)\tilde{\mathcal{N}}_{4^{'}}(\xi'){\delta }_{H_{4^{'}}}(\omega'))$ takes the largest value, and minimising the denominator with $12(\tilde{\mathcal{N}}_4(\xi){\delta }_{H_4}(\omega)N_{1^{'}}(\xi'){\delta }_{H_{1^{'}}}(\omega'))$ as this divides by the smallest of these terms in $f_2\left(\xi,\xi',\omega, \omega',l,l'\right)$. Using the normal distribution density, we expand these terms to obtain the following, again omitting the arguments in the distributions:
\begin{eqnarray}
\nonumber
{\cal{E}}_r \le
 \frac{1}{2}+\frac{3}{2}.\frac{4}{12}(\frac{\tilde{\mathcal{N}}_4{\delta }_{H_4}\tilde{\mathcal{N}}_{4^{'}}{\delta }_{H_{4^{'}}}}{\tilde{\mathcal{N}}_4{\delta }_{H_4}\tilde{\mathcal{N}}_{1^{'}}{\delta }_{H_{1^{'}}}}). 
\end{eqnarray}
 
\noindent We expand the Gaussian distribution where $(\tilde{\mathcal{N}}_4(\xi){\delta }_{H_4}(\omega)\tilde{\mathcal{N}}_{4^{'}}(\xi'){\delta }_{H_{4^{'}}}(\omega'))$ and  $(\tilde{\mathcal{N}}_4(\xi){\delta }_{H_4}(\omega)N_{1^{'}}(\xi'){\delta}_{H_{1^{'}}}(\omega'))$ are given by the following expressions respectively:

\begin{widetext}
\begin{equation}
\begin{gathered}
\nonumber
\tilde{\mathcal{N}}_4(\xi){\delta }_{H_4}(\omega)\tilde{\mathcal{N}}_{4^{'}}(\xi'){\delta }_{H_{4^{'}}}(\omega')
=e^-{\frac{{\xi-(J-\lambda(h_1+h_2))}^2}{2}e^-{\frac{{\xi'-(J-\lambda(h_1+h_2))}^2}{2}{\delta }_{H_4}{\delta }_{H_4'}}}\\
\tilde{\mathcal{N}}_4(\xi){\delta }_{H_4}(\omega)N_{1^{'}}(\xi'){\delta }_{H_{1^{'}}}(\omega')
=e^-{\frac{{\xi-(J-\lambda(h_1+h_2))}^2}{2}e^-{\frac{{\xi'-(J+\lambda(h_1+h_2))}^2}{2}{\delta }_{H_4}{\delta }_{H_1'}}}.
\end{gathered}
\end{equation}
\end{widetext}

\noindent Substituting these definitions into the bounds for ${\cal{E}}_r$ and by cancelling common terms, we find an upper bound, given by the following:
\begin{equation}
\begin{gathered}
{\cal{E}}_r \le \frac{1}{2}+\frac{1}{2}e^{2\lambda {\xi'}(h_1+h_2)}.
\end{gathered}
\end{equation}

\noindent Similarly, we bound from below through minimising the numerator using $\tilde{\mathcal{N}}_3(\xi){\delta }_{H_3}(\omega)\tilde{\mathcal{N}}_{3^{'}}(\xi'){\delta }_{H_{3^{'}}}(\omega')$ and maximising the denominator with $12(\tilde{\mathcal{N}}_3(\xi){\delta }_{H_3}(\omega)\tilde{\mathcal{N}}_{2^{'}}(\xi'){\delta }_{H_{2^{'}}}(\omega'))$ where we find the following, omitting the arguments in the distributions:
\begin{eqnarray}
\nonumber
{\cal{E}}_r \ge 
\frac{3}{2}.\frac{1}{12}(\frac{N_3{\delta }_{H_3}N_{3^{'}}{\delta }_{H_{3^{'}}}}{N_3{\delta }_{H_3}N_{2^{'}}{\delta }_{H_{2^{'}}}}). 
\end{eqnarray}

\noindent Once again, we expand Gaussian distributions with $\tilde{\mathcal{N}}_3(\xi){\delta }_{H_3}(\omega)\tilde{\mathcal{N}}_{3^{'}}(\xi'){\delta }_{H_{3^{'}}}(\omega')$ and $(\tilde{\mathcal{N}}_3(\xi){\delta }_{H_3}(\omega)\tilde{\mathcal{N}}_{2^{'}}(\xi'){\delta }_{H_{2^{'}}}(\omega'))$ given as follows:

\begin{widetext}
\begin{equation}
\begin{gathered}
\nonumber
\tilde{\mathcal{N}}_3(\xi){\delta }_{H_3}(\omega)\tilde{\mathcal{N}}_{3^{'}}(\xi'){\delta }_{H_{3^{'}}}(\omega')
=e^-{\frac{{\xi-(-J+\lambda(-h_1+h_2))}^2}{2}e^-{\frac{{\xi'-(-J+\lambda(-h_1+h_2))}^2}{2}{\delta }_{H_3}{\delta }_{H_3'}}}\\
\tilde{\mathcal{N}}_3(\xi){\delta }_{H_3}(\omega)N_{2^{'}}(\xi'){\delta }_{H_{2^{'}}}(\omega')
=e^-{\frac{{\xi-(-J+\lambda(-h_1+h_2))}^2}{2}e^-{\frac{{\xi'-(-J+\lambda(h_1-h_2))}^2}{2}{\delta }_{H_3}{\delta }_{H_2'}}}.
\end{gathered}
\end{equation}
\end{widetext}

\noindent Canceling common terms, we determine the lower bound such that we have the following bounds on ${\cal{E}}_r$ of the system:
\begin{equation}
\begin{gathered}
\frac{1}{8}e^{2\lambda {\xi'}(h_1-h_2)}\le {\cal{E}}_r \le \frac{1}{2}+\frac{1}{2}e^{2\lambda {\xi'}(h_1+h_2)}.
\end{gathered}
\end{equation}

\noindent Using these bounds, we find that the factorisation approximation does not hold, this is expected as in order to approximate a probability distribution function as a product of one-particle distribution functions the system must be factorisable. This clearly does not hold for the two qubit system as the terms are related via $J$. 

\section*{Appendix C: Derivation of the Evolution of Occupation Numbers }

\noindent In this section we establish the relationship shared between the occupation numbers and the level dynamics through the Pechukas model. Recall that a wave function on a Hilbert space can be expressed as the sum of linear combination of eigenstates that is:
\begin{equation}
\begin{gathered}
|\psi (t)\rangle=\sum_n{C_n(t)|n(t)\rangle}.
\end{gathered}
\end{equation}

\noindent For eigenstate coefficients for each fixed instant in time $C_n(t)\in \mathbb{C}$ , related to the occupation numbers $N_n$ by the following:
\begin{equation}
\begin{gathered}
{|C_n(t)|}^2=N_n.
\end{gathered}
\end{equation}

\noindent The evolution of $C_n$ associated to the eigenvalues of the state is obtained by:
\begin{eqnarray}
\nonumber
\mathcal{H}(t)|\psi (t)\rangle=i\frac{\partial }{\partial t}|\psi \rangle=\\
i\frac{\partial }{\partial t}\sum_n{C_n(t)|n(t)\rangle}=\sum_n{E_n(t)|\psi (t)\rangle}.
\end{eqnarray}

\noindent Taking time derivative: $.=\frac{\partial }{\partial t}$ using Leibniz rule, we obtain:
\begin{eqnarray}
\nonumber
i\frac{\partial |\psi \rangle}{\partial t}=
i\sum_n{\dot{C_n}(t)|n(t)\rangle+C_n(t)|\dot{n}(t)\rangle}=\\
\sum_n{C_n(t)E_n(t)|n(t)\rangle}.
\end{eqnarray}

\noindent Applying $\langle m(t)|$ on both sides and through linearity we obtain the dynamics of these coefficients through time with regards to the eigenvalues of the state. 
\begin{eqnarray}
\nonumber
i\sum_n{\dot{C_n}(t){\delta }_{mn}+{\langle m(t)|C}_n(t)|\dot{n}(t)\rangle}=\\
\sum_n{C_n(t)E_n(t){\delta }_{mn}}.
\end{eqnarray}

\noindent Hence by evaluating the $\delta $-distributions and rearranging the expression we have the following
\begin{eqnarray}
\nonumber
i\dot{C_m}(t)-C_m(t)E_m=\\
-i\sum_{n\neq m}{C_n(t)}\langle m(t)|\frac{\partial }{\partial \lambda }|n(t)\rangle \dot{\lambda }.
\end{eqnarray}

\noindent In order to evaluate these dynamics with respect to the level dynamics, it is necessary to determine $\sum_n{i}C_n(t)\langle m(t)|\frac{\partial }{\partial \lambda }|n\left(t\right)\rangle \dot{\lambda }$ where the term vanishes for $m(t)=n(t)$. Taking derivative with respect to $\lambda $ where terms are independent of coordinates other than time, we adopt the Pechukas model in order to express the evolution of $C_n(t)$ in terms of variables describing level dynamics. In determining the evolution of $\langle m(t)\left|\frac{\partial }{\partial \lambda }\right|n(t)\rangle$ we consider Eq. (1) such that we have the following relation:
 \begin{equation}
\begin{gathered}
\frac{\partial }{\partial \lambda }E_n(t)|n(t)\rangle\ =\ \frac{\partial }{\partial \lambda }\mathcal{H}(t)|n(t)\rangle.
\end{gathered}
\end{equation}

\noindent Applying the Leibniz rule on both sides we obtain:
\begin{eqnarray}
\nonumber
E_n(t)(\frac{\partial }{\partial \lambda }|n(t)\rangle)+|n(t)\rangle(\frac{\partial }{\partial \lambda }E_n(t))=\\
V(t)|n(t)\rangle+H(t)\frac{\partial }{\partial \lambda }|n(t)\rangle.
\end{eqnarray}

\noindent Acting on both sides with $\langle m(t)|$ and through linearity such that $m\neq n$, it reads:
 \begin{equation}
\begin{gathered}
E_n(t)\langle m(t)|\frac{\partial }{\partial \lambda }|n(t)\rangle=\\
\langle m(t)|V(t)|n(t)\rangle+E_m(t)\langle m(t)|\frac{\partial }{\partial \lambda }|n(t)\rangle.
\end{gathered}
\end{equation}

\noindent Hence 
 \begin{equation}
\begin{gathered}
(E_n(t)-E_m(t))\langle m(t)|\frac{\partial }{\partial \lambda }|n(t)\rangle\ =\langle m(t)|V(t)|n(t)\rangle.
\end{gathered}
\end{equation}

\noindent By applying the Pechukas equations to determine $l_{mn}$ as described in Eq. (4), we are able to determine the evolution of $\langle m(t)\left|\frac{\partial }{\partial \lambda }\right|n(t) \rangle$ entirely using level dynamics 
 \begin{equation}
\begin{gathered}
(x_n-x_m)\langle m(t)|\frac{\partial }{\partial \lambda }|n(t)\rangle\ =\frac{l_{mn}}{x_m-x_n}.
\end{gathered}
\end{equation}

\noindent * Thus 
 \begin{equation}
\begin{gathered}
<m(t)|\frac{\partial }{\partial \lambda }|n(t)>=\frac{-l_{mn}}{{(x_m-x_n)}^2}.
\end{gathered}
\end{equation}

\noindent Substituting this into Eq. (52) the dynamics of the occupation numbers are given through the relation:
 \begin{equation}
\begin{gathered}
i\dot{C_m}(t)-C_mx_m=i\dot{\lambda }\sum_{n\neq m}{C_n}\frac{l_{mn}}{{(x_m-x_n)}^2}.
\end{gathered}
\end{equation}

\noindent 
This is used to describe the wavefunction in its entirety at any given time.

* In the published paper, we had found there was a sign error in the equarion which carried through and has been corrected here. This has no implications on the conclusions of this paper establishing a link between the evolution of the quantum states and the level dynamics, simplifying to a reflection of initial conditions.

\end{document}